\documentclass[reprint,12pt,3p]{elsarticle}

\usepackage{amsmath,amssymb}
\usepackage{graphicx}               
\usepackage{wasysym}
\usepackage{lmodern}                
\usepackage[T1]{fontenc}            
\usepackage[utf8]{inputenc}         
\usepackage{indentfirst}            
\usepackage{physics}
\usepackage{csquotes}
\usepackage{cleveref}

\usepackage{natbib}
\begin{document}

\begin{frontmatter}

\title{From Hyperbolic to Non-Hyperbolic Open Billiards: An Entropy and Scaling Law Approach}

\author[UFPR]{P. Haerter\corref{teste}}%
\cortext[teste]{haerter@ufpr.br}
\author[UFSCAR]{A. F. Bosio}%
\author[UNESP]{E. D. Leonel}%
\author[URJC,RAEC]{M.A.F. Sanjuán}%
\author[UFPR,CICT]{R. L. Viana}%

\affiliation[UFPR]{organization={Departamento de Física, Universidade Federal do Paraná},
            city={Curitiba},
            postcode={81531-990}, 
            state={Paraná},
            country={Brazil}}
      
\affiliation[UFPR2]{organization={Centro Interdisciplinar de Ciência, Tecnologia e Inovação, Núcleo de Modelagem e Computação Científica, Universidade Federal do Paraná},
            city={Curitiba},
            postcode={81531-990}, 
            state={Paraná},
            country={Brazil}}

\affiliation[UFSCAR]{organization={Department of Computer Science, Federal University of São Carlos},
            city={São Carlos},
            postcode={13565-905}, 
            state={São Paulo},
            country={Brazil}}

\affiliation[UNESP]{organization={Departamento de Física, Universidade Estadual Paulista},
            city={Rio Claro},
            postcode={13506-900}, 
            state={São Paulo},
            country={Brazil}}

\affiliation[URJC]{organization={Nonlinear Dynamics, Chaos, and Complex Systems Group, Departamento de Física, Universidad Rey Juan Carlos},
            city={Móstoles},
            postcode={28933}, 
            state={Madrid},
            country={Spain}}

\affiliation[RAEC]{organization={Royal Academy of Sciences of Spain},
            postcode={28004}, 
            state={Madrid},
            country={Spain}}

\begin{abstract}
We investigate the escape dynamics in an open circular billiard under the influence of a uniform gravitational field. The system properties are investigated as a function of the particle total energy and the size of two symmetrically placed holes in the boundary. Using a suite of quantitative tools including escape basins, basin entropy ($S_b$), mean escape time ($\bar{\tau}$), and survival probability ($P(n)$), we characterize a system that transitions from a fully chaotic, hyperbolic regime at low energies to a non-hyperbolic, mixed phase space at higher energies. Our results demonstrate that this transition is marked by the emergence of Kolmogorov-Arnold-Moser (KAM) islands. We show that both the basin entropy and the mean escape time are sensitive to this transition, with the former peaking and the latter increasing sharply as the sticky KAM islands appear. The survival probability analysis confirms this dynamical picture, shifting from a pure exponential decay in the hyperbolic regime to a power-law-like decay with a saturation plateau in the mixed regime, which directly quantifies the measure of trapped orbits. In the high-energy limit, the system dynamics approaches an integrable case, leading to a corresponding decrease in complexity as measured by both $S_b$ and $\bar{\tau}$.
\end{abstract}


\end{frontmatter}

\section{Introduction}

Billiard systems are among the most studied paradigms in nonlinear dynamics and statistical mechanics, serving as powerful yet conceptually simple models for exploring the foundations of chaos in Hamiltonian systems\cite{birkhoffDynamicalSystems1966,chernovChaoticBilliards2006}. The dynamics of a point particle moving freely within a closed boundary and undergoing specular reflections can range from fully integrable, as in a rectangular or circular billiard, to fully chaotic and ergodic, as in the Sinai billiard\cite{sinaiDynamicalSystemsElastic1970}. The study of these systems provides fundamental insights into concepts such as ergodicity, mixing, and the transition to chaos\cite{arnoldErgodicProblemsClassical1989,borgonoviDiffusionLocalizationChaotic1996,firmbachThreedimensionalBilliardsVisualization2018,dullinLinearStabilityBilliards1998}.

A particularly rich area of investigation arises when these systems are made "leaky" by introducing one or more holes in the boundary, allowing particles to escape\cite{rolimsalesInvestigationEscapeScaling2024,sanjuanOpeningClosedHamiltonian2003,schneiderDynamicsLeakingHamiltonian2002a}. Such open systems provide a framework for understanding transient chaos and chaotic scattering\cite{altmannLeakingChaoticSystems2013,afraimovichWhichHoleLeaking2010,altmannPoincareRecurrencesTransient2009}. The set of initial conditions that lead to escape through a specific exit forms a structure known as an escape basin. The boundaries of these basins are often fractal, revealing the intricate and sensitive dependence on initial conditions that is the hallmark of chaos\cite{aguirreFractalStructuresNonlinear2009,ottStrangeAttractorsChaotic1981,poonWadaBasinBoundaries1996,percivalChaoticBoundaryHamiltonial1982}.

Further complexity can be introduced by subjecting the billiard to an external field, which breaks the symmetries of the free system\cite{robnikClassicalBilliardsMagnetic1985,loskutovPropertiesChaoticBilliards2000,deryabinGeneralizedRelativisticBilliards2003,lehtihetNumericalStudyBilliard1986}. In this work, we investigate a circular billiard under the influence of a uniform gravitational field. The addition of gravity changes particle trajectories between collisions from straight lines to parabolic arcs, profoundly altering the phase space structure\cite{dacostaCircularEllipticOval2015}. By introducing two holes on the boundary of this gravitational billiard, we create an open system where the escape dynamics depends critically on the interplay between the boundary geometry, the gravitational field, and the particle total energy.

The primary goal of this paper is to conduct a detailed numerical investigation into how the escape properties and the structure of the phase space of the gravitational open billiard change as a function of two key parameters: the total energy of the particle, and the size of the escape holes. To quantify these changes, we compute the basin entropy, $S_b$, a powerful measure of the uncertainty and unpredictability of the final state\cite{dazaBasinEntropyNew2016,dazaUnpredictabilityBasinEntropy2023, dazaMultistabilityUnpredictability2024}. We complement this spatial analysis with a temporal one, by computing the mean escape time,  and the survival probability of the particles\cite{rolimsalesInvestigationEscapeScaling2024}.

Our results reveal a distinct transition from a strongly chaotic hyperbolic regime at low energies to a non-hyperbolic, mixed phase space at higher energies. This transition, marked by the emergence of large Kolmogorov-Arnold-Moser (KAM) islands, is clearly detected by both the basin entropy and the mean escape time, demonstrating their effectiveness as indicators of changes in the underlying dynamics. Furthermore, we show that at very high energies, the system dynamics approaches an integrable limit, leading to a decrease in both chaos and escape time. The survival probability analysis confirms this picture by showing how the decay laws change from purely exponential in the fully chaotic regime to a power-law-like decay with permanent trapping in the presence of KAM islands.

This paper is organized as follows. In Section II, we describe the model in detail and derive the mapping equations that govern the dynamics. Section III is dedicated to the analysis of the escape basins and their characterization using basin entropy. In Section IV, we discuss the temporal characteristics of the system through the mean escape time and survival probability analysis. Finally, Section V presents our conclusions.

\section{The Model and the mapping}

\begin{figure}[htp]
    \centering
    \includegraphics[width=0.6\textwidth]{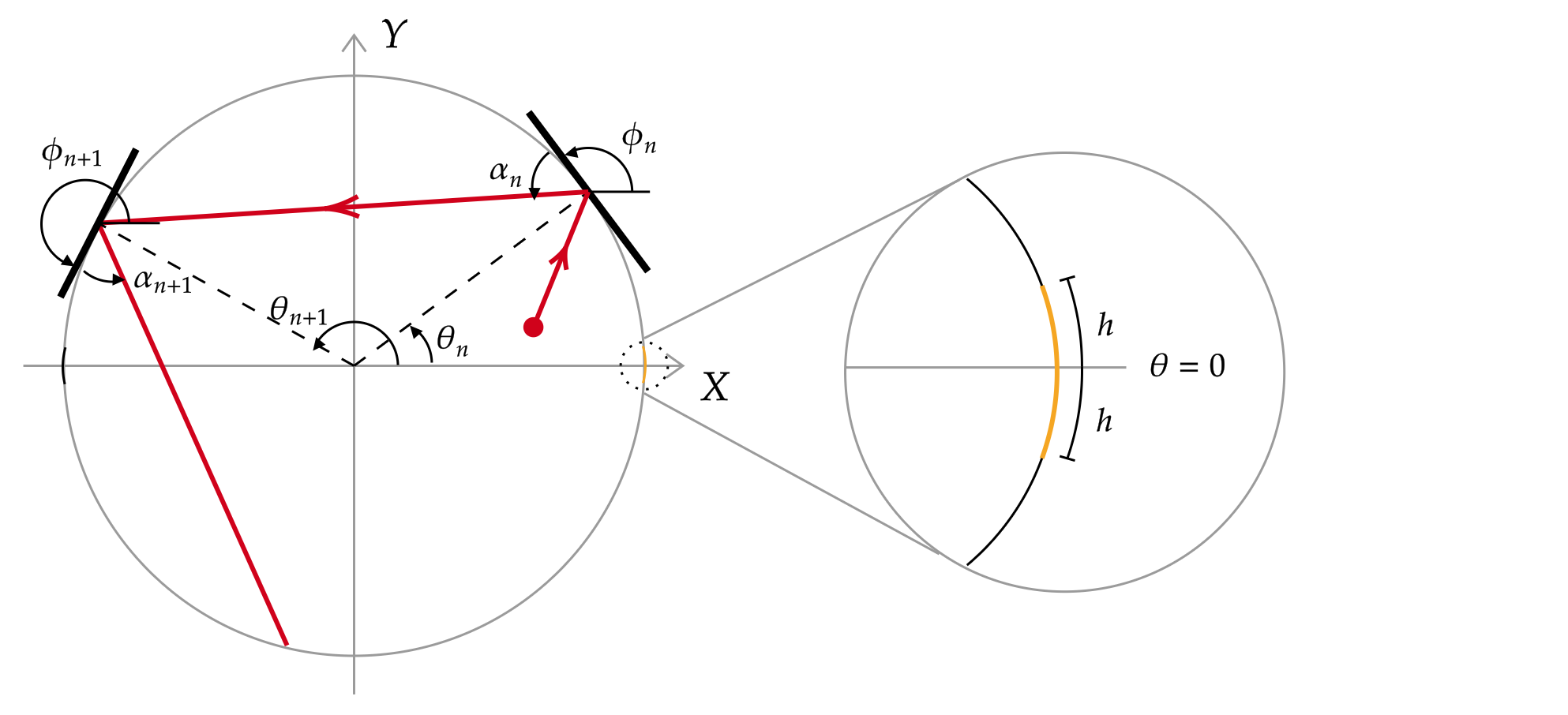}
    \caption{Schematic of the open circular billiard under a uniform gravitational field. The diagram illustrates a particle parabolic trajectory between two collisions, defining the angular position $\theta_n$ and the reflection angle $\alpha_n$. The angle of the tangent at the collision point is $\phi_n$. The inset shows a magnified view of one of the two holes, each with an angular width of $2h$.}
    \label{fig:diagrama}
\end{figure}

The billiard model consists of a classical point particle of mass $m$ undergoing elastic collisions with a static, circular boundary of unit radius, $R_b = 1$, this model is already present in the literature, and it is known for its rich dynamics \cite{dacostaCircularEllipticOval2015}. The dynamics of the model is described by a four-dimensional map, $T(\theta_n, \alpha_n, V_n, t_n) = (\theta_{n+1}, \alpha_{n+1}, V_{n+1}, t_{n+1})$. Here, the variables denote the angular position of the particle ($\theta_n$), the angle between the particle trajectory and the tangent to the boundary at the collision point ($\alpha_n$), the particle velocity ($V_n$), and the time of the collision ($t_n$), respectively. Figure~\ref{fig:diagrama} illustrates the geometry of three successive collisions of the particle with the boundary.

From the four-dimensional map, the particle collision with the boundary is characterized by the two angles $\theta$ and $\alpha$. The two initial coordinates that describe the position of the particle are
\begin{align}
    X(\theta_n) = R_b\cos(\theta_n),\\
    Y(\theta_n) = R_b\sin(\theta_n).
\end{align}

It is also convenient to define the auxiliary angle $\phi_n$ as the slope of the tangent line at the point of collision, measured counterclockwise from the horizontal axis. This angle is defined as
\begin{equation}
    \phi_n = \arctan\qty(\frac{Y'(\theta_n)}{X'(\theta_n)}) \pmod{2\pi},
\end{equation}
where the prime denotes the derivative with respect to $\theta$. Therefore, the direction of the particle momentum vector, $\eta_n$, is given by
\begin{equation}
    \eta_n = \phi_n + \alpha_n \pmod{2\pi}.
\end{equation}

This allows us to write the particle velocity vector, $\vb{V}_n$, as
\begin{equation}
    \vb{V}_n = \abs{\vb{V}_n}\qty[\cos(\eta_n)\vu{i} + \sin(\eta_n)\vu{j}],
\end{equation}
where $\vu{i}$ and $\vu{j}$ are the Cartesian unit vectors.

Since a gravitational force acts on the particle during its flight, the trajectory between collisions consists of parabolic arcs. For a time $t \geq t_n$, the particle position, $(X_p(t), Y_p(t))$, is given by
\begin{align}
    X_p(t) &= X(\theta_n) + \abs{\vb{V}_n}\cos(\eta_n)\Delta t, \\
    Y_p(t) &= Y(\theta_n) + \abs{\vb{V}_n}\sin(\eta_n)\Delta t - \frac{1}{2}g\qty(\Delta t)^2,
\end{align}
where $\Delta t = t - t_n$ is the time elapsed since the $n$-th collision and  $g$ is the gravitational acceleration. The particle distance from the origin at time $t$ is then
\begin{equation}
    R_p(t) = \sqrt{X_p^2(t)+Y_p^2(t)}.
\end{equation}
Therefore, the angular position of the next collision, $\theta_{n+1}$, is found by solving for the time $t_{n+1}$ that satisfies the condition
\begin{equation}
    R_p(t_{n+1}) = R_b = 1.
\end{equation}

The Cartesian components of the velocity vector just before the $(n+1)$-th collision, $(V_x, V_y)$, are determined by the state at the $n$-th collision and the time of flight, $\Delta t$, so that
\begin{align}
    V_x &= V_n\cos(\eta_n), \\
    V_y &= V_n\sin(\eta_n) - g\Delta t.
\end{align}
For an elastic collision with a static boundary, the velocity component tangent to the boundary is conserved, while the normal component is inverted. We can define the tangential ($V_T$) and normal ($V_N$) components of the velocity vector after the collision in terms of the pre-collision components $(V_x, V_y)$ and the tangent angle $\phi_{n+1}$ is given by
\begin{align}
    V_T &= V_x\cos(\phi_{n+1}) + V_y\sin(\phi_{n+1}), \label{eq:VT}\\
    V_N &= -\qty(V_x\sin(\phi_{n+1}) - V_y\cos(\phi_{n+1})). \label{eq:VN}
\end{align}
The magnitude of the particle velocity after the collision, $V_{n+1}$, is the magnitude of this new velocity vector:
\begin{equation}
    V_{n+1} = \sqrt{V_T^2 + V_N^2}.
\end{equation}
Finally, the new reflection angle, $\alpha_{n+1}$, is defined as the angle between the outgoing velocity and the tangent
\begin{equation}
    \alpha_{n+1} = \arctan\qty(\frac{V_N}{V_T}).
\end{equation}

\begin{figure*}[ht!]
    \centering
    \includegraphics[width=0.8\textwidth]{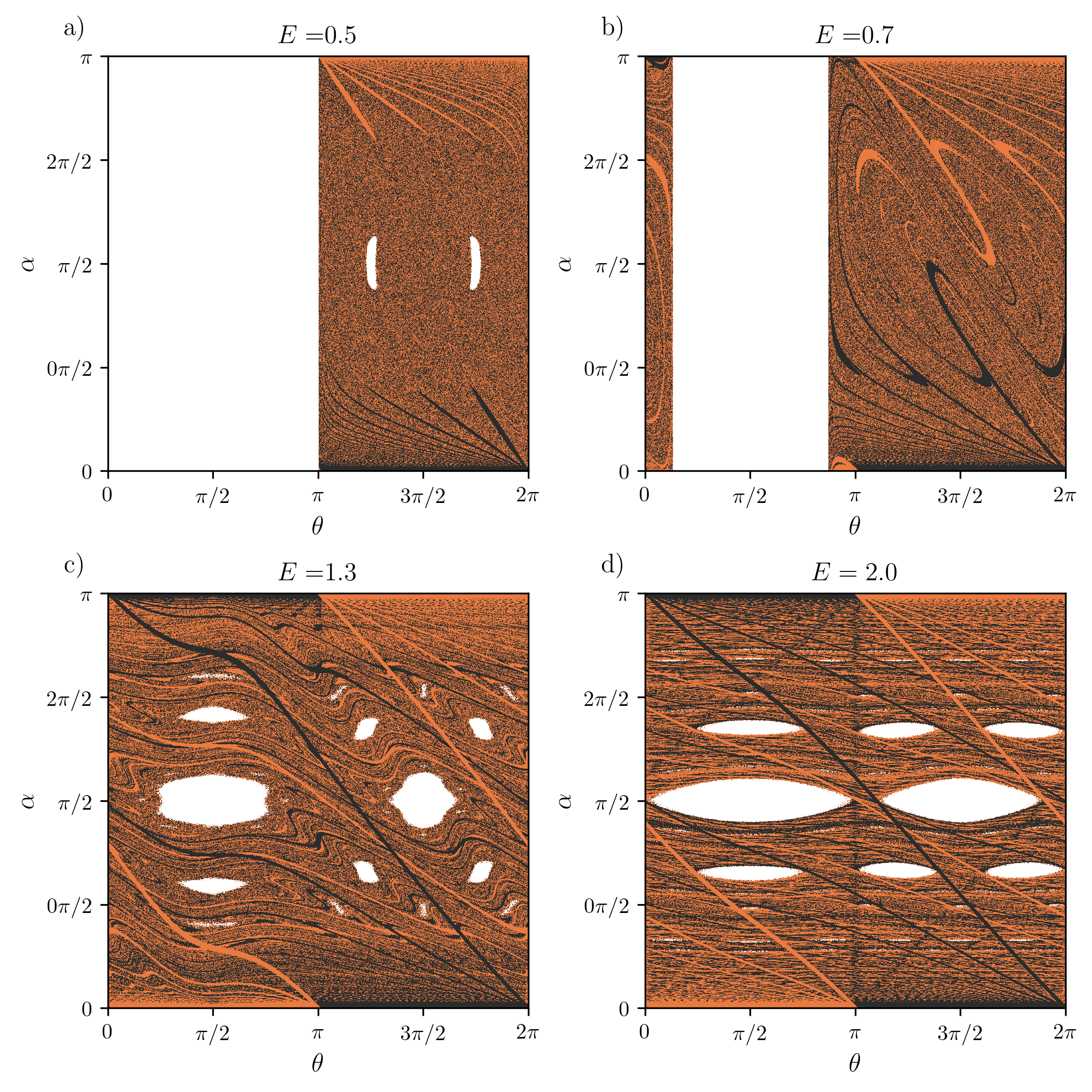}
    \caption{Escape basins in the $(\theta, \alpha)$ phase space for a fixed hole size of $h=0.04$ at four characteristic total energies: (a) $E=0.5$ and (b) $E=0.7$ in the hyperbolic-like regime, and (c) $E=1.3$ and (d) $E=2.0$ in the mixed phase space regime. Initial conditions that escape through the hole at $\theta=0$ are colored orange, while those escaping at $\theta=\pi$ are black. White regions correspond to initial conditions that do not escape within $10^6$ collisions, indicating the presence of KAM islands.}
    
    \label{fig:big_Basins}
\end{figure*}

To obtain the system orbits in phase space, we consider particles with a fixed total energy $E$, which is conserved during the flight between collisions. The total energy is the sum of the kinetic energy, $K=V^2/2$, and the gravitational potential energy, $U=gh$. 

We set the reference level for potential energy ($U=0$) at the bottom of the billiard, which corresponds to a vertical position of $Y=-1$. The particle height above this level is thus $h=Y+1$. Since the particle is on a circular boundary of unit radius, its vertical position is $Y(\theta)=\sin(\theta)$. The potential energy is therefore a function of the angular position:
\begin{equation}
    U(\theta) = g(\sin(\theta)+1).
\end{equation}
For a given total energy $E$, the initial speed $V_0$ of a particle starting at an angle $\theta_0$ is determined by the energy conservation as, $E = K_0 + U(\theta_0)$, and is given by
\begin{equation}
    V_0 = \sqrt{2\qty(E - U(\theta_0))}.
\end{equation}
For this initial speed to be a real number, the total energy must be greater than or equal to the potential energy at the starting position, i.e., $E \ge U(\theta_0)$ \cite{dacostaCircularEllipticOval2015}.

 \section{Escape Basins}

In this section, we investigate the escape dynamics of particles from the billiard through two holes on its boundary. Each hole is defined by an angular width of $2h$. For the numerical simulations, we set the gravitational acceleration to $g=0.5$ and the mass as $m=1$, and we study the system behavior for several different values of the total energy, $E$. 

The two holes are centered on the horizontal axis at angular positions $\theta_{\text{exit}}^{(1)}=0$ and $\theta_{\text{exit}}^{(2)}=\pi$. This symmetric placement is chosen to avoid any intrinsic bias in the escape direction. Due to the system up-down symmetry, the probability of a particle escaping through either hole is, a priori, the same.

\begin{figure*}[ht]
    \centering
    \includegraphics[width=0.8\textwidth]{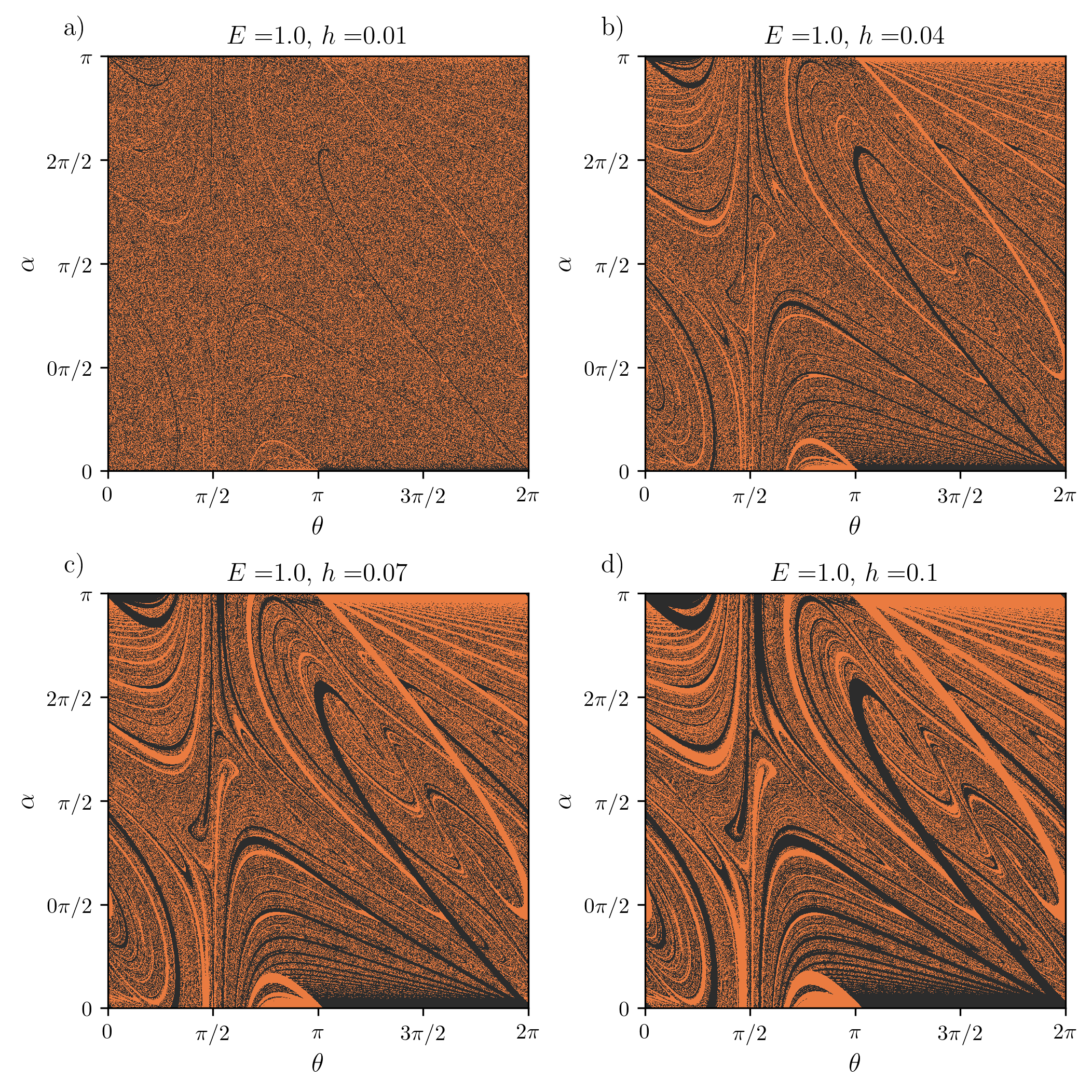}
    \caption{Escape basins for the open billiard at a fixed total energy of $E=1.0$ for four different hole sizes: (a) $h=0.01$, (b) $h=0.04$, (c) $h=0.07$, and (d) $h=0.1$. The color scheme is the same as in Fig.~\ref{fig:big_Basins}. As the hole size increases, the basin structures become larger and more defined, reflecting a decrease in the system unpredictability.}
   
    \label{fig:big_h}
\end{figure*}

We initialize an ensemble of $M=2^{20}$ particles, which are distributed uniformly in the phase space region defined by $(\theta, \alpha) \in [0, 2\pi] \times [0, \pi]$. To construct the escape basins, we iterate the dynamics of each particle until it escapes through one of the two holes or reaches a maximum of $N=10^6$ collisions. If a particle escapes through hole \#1 (centered at $\theta=0$) or hole \#2 (centered at $\theta=\pi$), we color its initial condition in the phase space orange or black, respectively. If a particle does not escape within the maximum number of iterations, its initial condition is colored white.

Figure~\ref{fig:big_Basins} shows the escape basins for four different values of the total energy $E$, all for a fixed hole size of $h=0.04$. The chosen energy values include two values below the critical energy ($E \approx 1$), one in an intermediate regime with a mixed phase space, and one at $E=2$, where the system dynamics begin to resemble the integrable case. The effect of varying the hole size $h$ is primarily a change in the overall size of the basins and the absorption of uncertain regions into larger, smoother basins. This is illustrated in Fig.~\ref{fig:big_h}, which shows a comparison of the basins for four different values of $h$.

In order to quantify the properties of different configurations, we apply the concept of basin entropy \cite{dazaBasinEntropyNew2016}. This method has been successfully applied to a variety of problems in nonlinear dynamics, such as neuronal analysis, celestial mechanics, applied systems and magnetic systems\cite{nietoMeasuringTransitionNonhyperbolic2020,leAsymmetricCouplingNonchaotic2025,zotosBasinsAttractionEquilibrium2018,halekotteTransientChaosEnforces2021,jensenFullPhaseSpace2021}.

To quantify the effects of different final states, we employ basin entropy analysis\cite{dazaBasinEntropyNew2016}. This method partitions a region of interest $\Omega$ in the phase space into a grid of $N$ boxes, each with a linear size of $\varepsilon$. For a given set of parameters, we consider a set of $N_A$ possible final states, often referred to as ``colors.''

For each box $i$, we can determine the probability, or observed frequency, $p_{i,j}$ associated with each color $j$. This is calculated by launching $N_{ic}$ initial conditions from within the box and finding the fraction that leads to a specific color
\begin{equation}
    p_{i,j} = \frac{n_j}{N_{ic}},
\end{equation}
where $n_j$ is the number of initial conditions in box $i$ that result in color $j$.

Assuming the trajectories within a box are statistically independent, the Gibbs entropy for each box $i$ is
\begin{equation}
    S_i = -\sum_{j=1}^{m_i} p_{i,j} \log p_{i,j},
\end{equation}
where $m_i$ is the number of distinct colors actually found within box $i$ ($m_i \in [1, N_A]$).

To create a metric that is independent of the grid resolution, the basin entropy, $S_b$, is defined as the average entropy per box
\begin{equation}
    S_b = \frac{1}{N} \sum_{i=1}^{N}S_i.
\end{equation}

This metric quantifies the overall unpredictability of the system. A value of $S_b \to 0$ indicates that most initial conditions lead to a single, predictable outcome. Conversely, a value of $S_b \to \log N_A$ corresponds to maximal uncertainty, where outcomes are randomly distributed. The basin entropy is closely related to classical metrics like the fractal dimension and the uncertainty exponent\cite{dazaBasinEntropyNew2016}.

In our case, there are three possible asymptotic states for a particle: it can escape through hole \#1, escape through hole \#2, or remain confined within the billiard. Therefore, the number of possible outcomes is $N_A=3$. To calculate the basin entropy, we first determine the escape basins for various total energies in the range $E \in [1.0, 2.0]$ and for multiple hole sizes in the interval $h \in [0.005, 0.1]$.

\begin{figure*}[ht]
    \centering
    \includegraphics[width=0.7\textwidth]{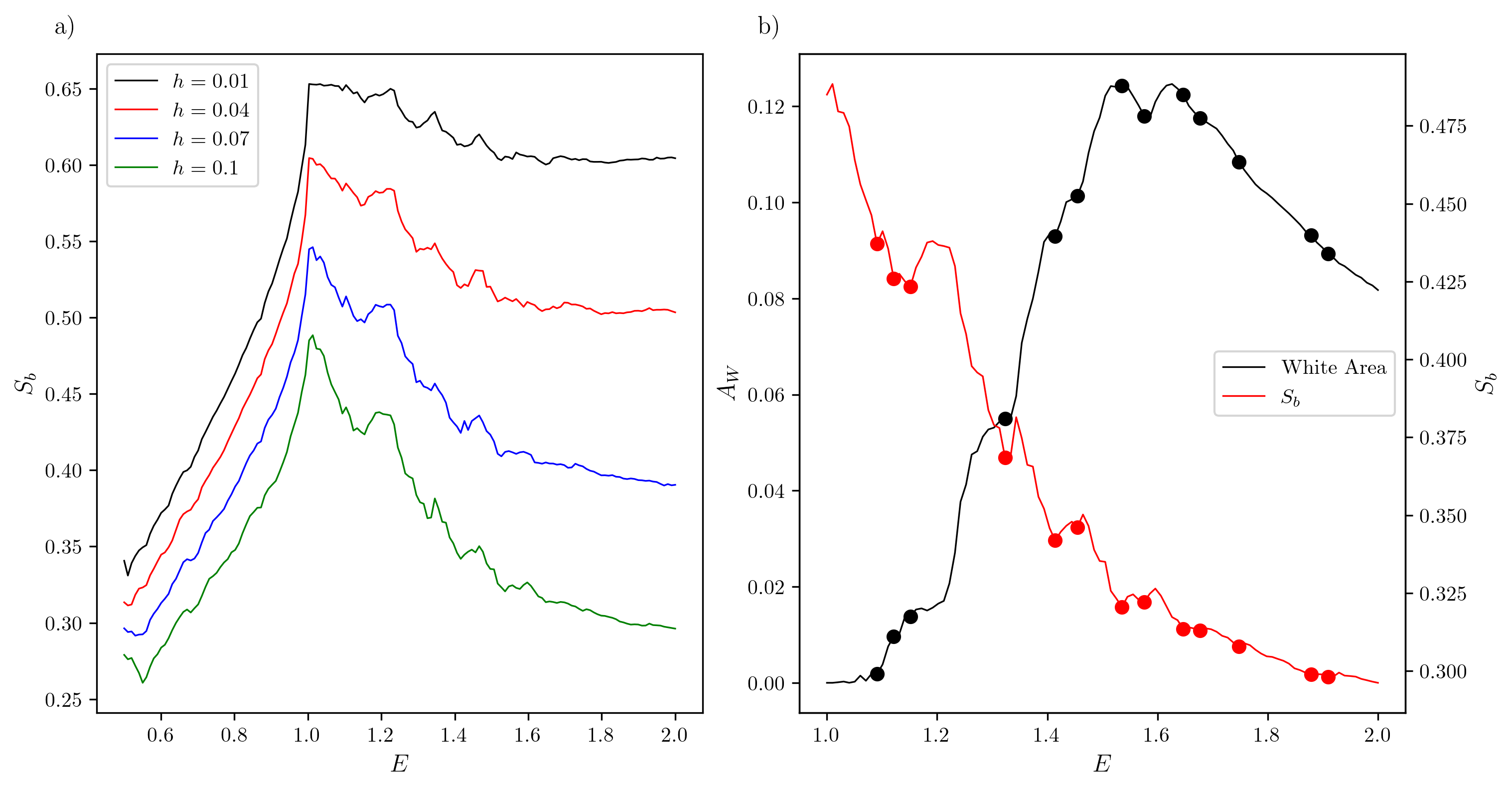}
   \caption{(a) Basin entropy $S_b$ as a function of the total energy $E$ for four different hole sizes, as indicated in the legend. A peak in $S_b$ is observed near the transition energy $E \approx 1.0$. (b) A comparison between the phase space area of non-escaping orbits ($A_w$, black curve, left axis) and the basin entropy ($S_b$, red curve, right axis) for a fixed hole size of $h=0.1$. The local minima of $S_b$ correlate with sharp increases in the area of the KAM islands.}
    \label{fig:S_b}
\end{figure*}

Figure~\ref{fig:S_b}(a) shows the results of this analysis. For small values of $h$, the basin entropy $S_b$ is large, which is expected as the escape basins exhibit a fine, highly intermingled structure with minimal large-scale order. As the hole size $h$ increases, coherent structures begin to emerge in the phase space and $S_b$ generally decreases. This decrease occurs because larger holes create larger, smoother basin regions that absorb the previously uncertain, fractal-like boundaries, thus reducing the overall unpredictability of the system.

For each selected value of the hole size $h$, the basin entropy $S_b$ exhibits a similar dependence on the total energy $E$, as shown in Fig.~\ref{fig:S_b}(a). Two distinct behaviors are observed: for $E \lesssim 1.0$, the basin entropy increases monotonically, while for $E \gtrsim 1.0$, $S_b$ fluctuates. The first region corresponds to a strongly chaotic, or hyperbolic, regime. This agrees with previous results which demonstrated that for lower energies, the system exhibits apparent ergodicity, a characteristic of hyperbolic systems \cite{dacostaCircularEllipticOval2015}. The second, fluctuating region corresponds to the emergence of a mixed phase space, signaling a transition to a non-hyperbolic regime. Our numerical analysis of the phase space supports these findings, confirming the appearance of large KAM islands for $E \gtrsim 1.0$.

This change in the behavior of $S_b$ can be directly related to the system transition from a hyperbolic to a non-hyperbolic regime. As the energy $E$ increases, the system undergoes a bifurcation. For values of $E > 2gR_b$ (with $R_b=1$), particles can reach the top of the billiard, which allows for the existence of new stable periodic orbits and, consequently, the generation of KAM islands \cite{dacostaCircularEllipticOval2015}. While this transition was originally identified using other methods, our analysis shows that the basin entropy $S_b$ is also a sensitive indicator of this change in dynamics. 

This is further evidenced by comparing the basin entropy to the phase space area occupied by non-escaping initial conditions (i.e., those within KAM islands), as shown in Fig.~\ref{fig:S_b}(b). The comparison reveals that the local minima of $S_b$ correspond to the abrupt increases in the size of the KAM islands, a result consistent with findings in similar systems \cite{nietoMeasuringTransitionNonhyperbolic2020}.

\section{Escape Time and Survival Probabillity}

In addition to the final destination of the particles, we also analyze the time required for them to escape. By calculating the mean escape time, $\bar{\tau}$, we can observe the same transition in the system characteristics that was identified by the basin entropy. The behavior of $\bar{\tau}$ as a function of energy also reveals two distinct regimes.

\begin{figure}[ht]
    \centering
    \includegraphics[scale=0.4]{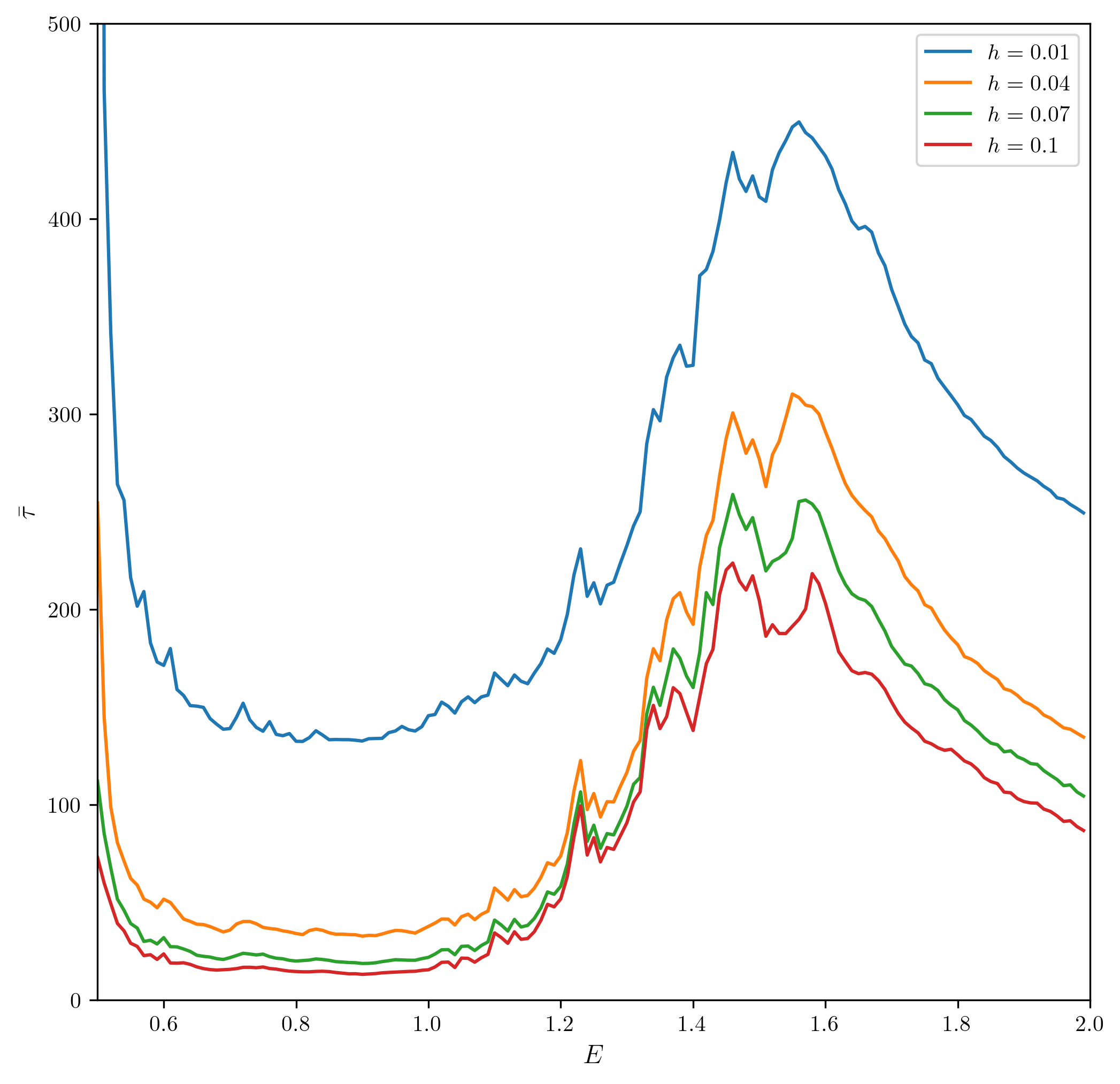}
    \caption{Mean escape time $\overline{\tau}$ as a function of total energy $E$ for four different hole sizes $h$. The sharp increase in $\overline{\tau}$ for $E \ge 1.0$ is a direct consequence of the 'stickiness' effect introduced by the emergence of KAM islands in the phase space.}
   
    \label{fig:tau}
\end{figure}

As shown in Fig.~\ref{fig:tau}, for lower energies corresponding to the hyperbolic regime ($E \lesssim 1.0$), particles tend to escape relatively quickly. However, as the energy approaches and surpasses the critical value of $E \approx 1.0$, the mean escape time begins to increase significantly. This increase is due to the presence of KAM islands that emerge within the chaotic sea. These islands act as "sticky" regions, temporarily trapping trajectories that start near their boundaries and preventing them from reaching an exit, which in turn raises the average time a particle spends inside the billiard before escaping\cite{custodioIntrinsicStickinessChaos2011,contopoulosSTICKINESSCHAOS2008,zouCharacterizationStickinessMeans2007}.

For energies exceeding $E > 1.6$, we observe a simultaneous decrease in both the mean escape time $\bar{\tau}$ and the basin entropy $S_b$. This behavior can be attributed to the system approaching its integrable limit. At high energies, the kinetic energy of the particle is significantly larger than the potential energy variations caused by gravity. As a result, the relative effect of the gravitational field becomes negligible, and the system dynamics begin to approximate those of a standard, integrable circular billiard \cite{dacostaCircularEllipticOval2015}.

In this high-energy limit ($E \to \infty$), the complex, mixed phase space structure with KAM islands and a chaotic sea dissolves. Trajectories become regular across the billiard. This significant reduction in chaotic behavior leads to less uncertainty about the final destination of a particle, thus lowering the basin entropy $S_b$. Concurrently, the predictability of the paths allows for more direct and rapid escape routes, which decreases the mean escape time $\bar{\tau}$.

We also compute the survival probability, $P(n)$, which is the fraction of particles that have not escaped through a hole up to the $n$-th collision. This is defined as
\begin{equation}
    P(n) = \frac{N_s(n)}{M},
\end{equation}
where $N_s(n)$ is the number of particles that have survived at least $n$ collisions, and $M$ is the total number of initial conditions in the ensemble. The behavior of $P(n)$ varies depending on the system dynamical regime, which is controlled by the energy $E$. For systems with a mixed phase space, the decay rate is slow, often resembling a power law or a stretched exponential. This slow decay is caused by "stickiness" regions surrounding KAM islands, where particles can be trapped for long, but finite, times before escaping. In contrast, for strongly chaotic (hyperbolic) systems, the survival probability typically decays exponentially.

\begin{figure}[tb]
     \centering
     \includegraphics[scale=0.5]{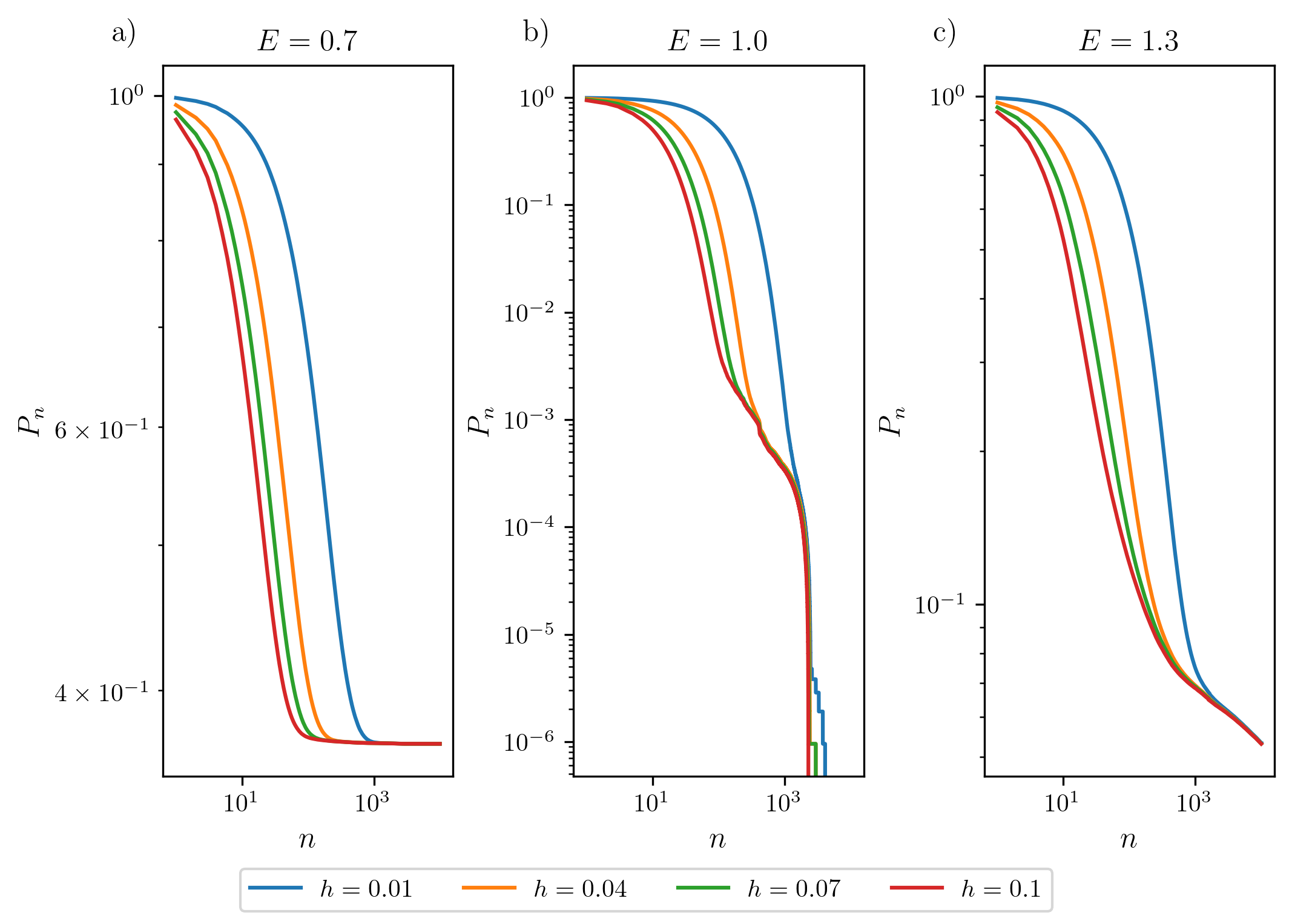}
     \caption{Survival probability $P(n)$ as a function of the number of collisions $n$ for several hole sizes $h$. The three panels correspond to three characteristic energy regimes: (a) a mixed phase space at $E=0.7$, (b) the transitional hyperbolic regime at $E=1.0$, and (c) a mixed phase space with larger islands at $E=1.3$. Note the saturation for large $n$ in panels (a) and (c), which corresponds to the fraction of permanently trapped orbits.}
     \label{fig:kappa}
 \end{figure}

In Fig.~\ref{fig:kappa}, we present the survival probability for four different hole sizes, $h$, calculated for three characteristic energy scenarios previously shown in Figs.~\ref{fig:big_Basins} and \ref{fig:big_h}: a hyperbolic case ($E=0.7$), a transitional case where large sticky regions are absent ($E=1.0$), and a mixed phase space case ($E=1.3$).

The cases with a mixed phase space ($E=0.7$ and $E=1.3$) both exhibit a decay resembling a stretched exponential, followed by a saturation of the survival probability at a constant value for large $n$. This saturation plateau represents the fraction of particles located on KAM islands, which are permanently trapped and will never escape. However, when these KAM islands are effectively removed by setting the energy to the transitional value of $E=1.0$, the survival probability behaves as a pure exponential decay. This confirms that, in the absence of stable islands, most particles will eventually escape the system.

To further probe the system's dynamics and search for universal properties, we investigate the scaling behavior of the survival probability with respect to the hole size, $h$. Scaling laws are a powerful tool in nonlinear dynamics, as their existence implies that the system's behavior is self-afine and governed by a consistent underlying principle, regardless of the specific parameter scale. Our goal is to demonstrate that the family of decay curves for different hole sizes can be collapsed onto a single, universal curve through an appropriate rescaling of the time axis, a technique successfully employed in similar billiard systems \cite{hansenInfluenceStabilityIslands2016,rolimsalesInvestigationEscapeScaling2024}.

\begin{figure}[hb]
    \centering
    \includegraphics[scale=0.5]{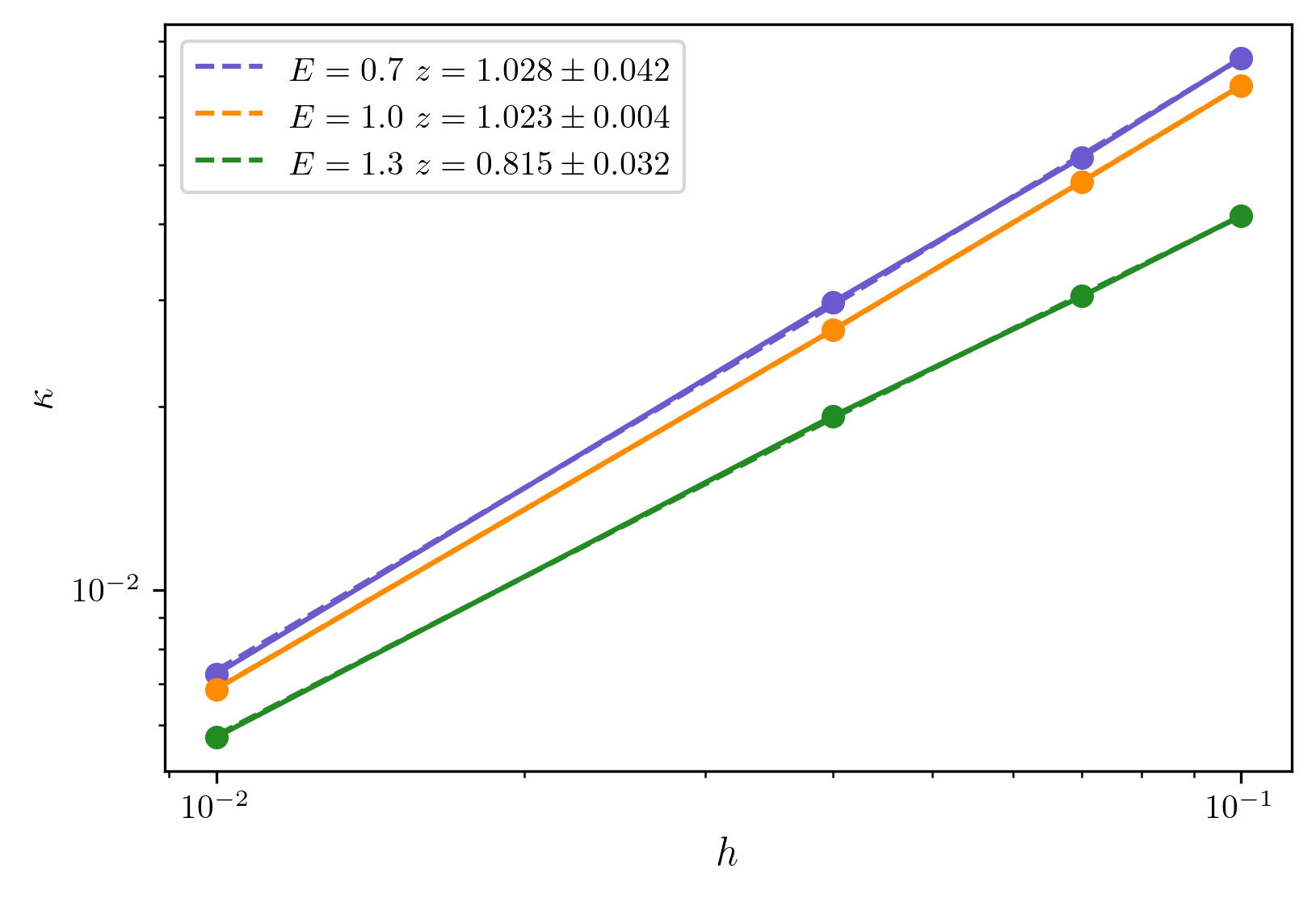}
    \caption{The escape rate $\kappa$ for $E = 0.7$ (blue dots), $E = 1.0$ (orange dots) and $E=1.3$ (green dots) as a function of the hole size h. The dashed lines correspond to the optimal fit based on the function $f(h)\sim h^z$}
    \label{fig:zfit}
\end{figure}

Our analysis begins with the dynamical regimes that exhibit permanent particle trapping (namely for energies $E < 1.0$ and $E > 1.6$), where a fraction of orbits never escape. As established, the survival probability $P(n)$ in these cases is well-described by a exponential function with a non-zero offset:
\begin{equation}
    P(n) \propto e^{-\kappa n } +\Gamma
    \label{eq:stretched_offset}
\end{equation}
Here, $\Gamma$ represents the fraction of non-escaping orbits, ranging from $\Gamma \in [0,1]$. A value of $\Gamma=1$ indicates that no particles escape, while $\Gamma=0$ means that all particles escape. The parameter $\kappa$ can be interpreted as an effective escape rate for particles in the chaotic sea. By fitting this equation to the data for each hole size $h$, we found that the escape rate $\kappa$ shows a clear dependence on $h$. A log-log plot of $\kappa$ versus $h$ reveals a power-law relationship, $\kappa \propto h^z$, as shown in Fig.~\ref{fig:zfit}. In this relationship, $z$ is the scaling exponent that characterizes the system's response to the leak size, with measured values of $z_{0.7}=1.028\pm0.0042$ and $z_{1.3}=0.815\pm0.032$.

A distinct analysis was required for the transitional regime around $E=1.0$, where the KAM islands vanish and all particles eventually escape. In this case, the survival probability follows a pure exponential decay of the form
\begin{equation}
    P(n) \propto e^{-\kappa n }.
    \label{eq:stretched_pure}
\end{equation}
Remarkably, the escape rate $\kappa_1$ for this hyperbolic-like regime was also found to obey a power-law dependence on the hole size, $\kappa_1 \propto h^{z_1}$, albeit with a different scaling exponent $z_1=1.023\pm0.004$. This finding suggests that scaling invariance is a robust feature of the system, present in both its hyperbolic and non-hyperbolic configurations.

\begin{figure}[tb]
    \centering
    \includegraphics[scale=0.5]{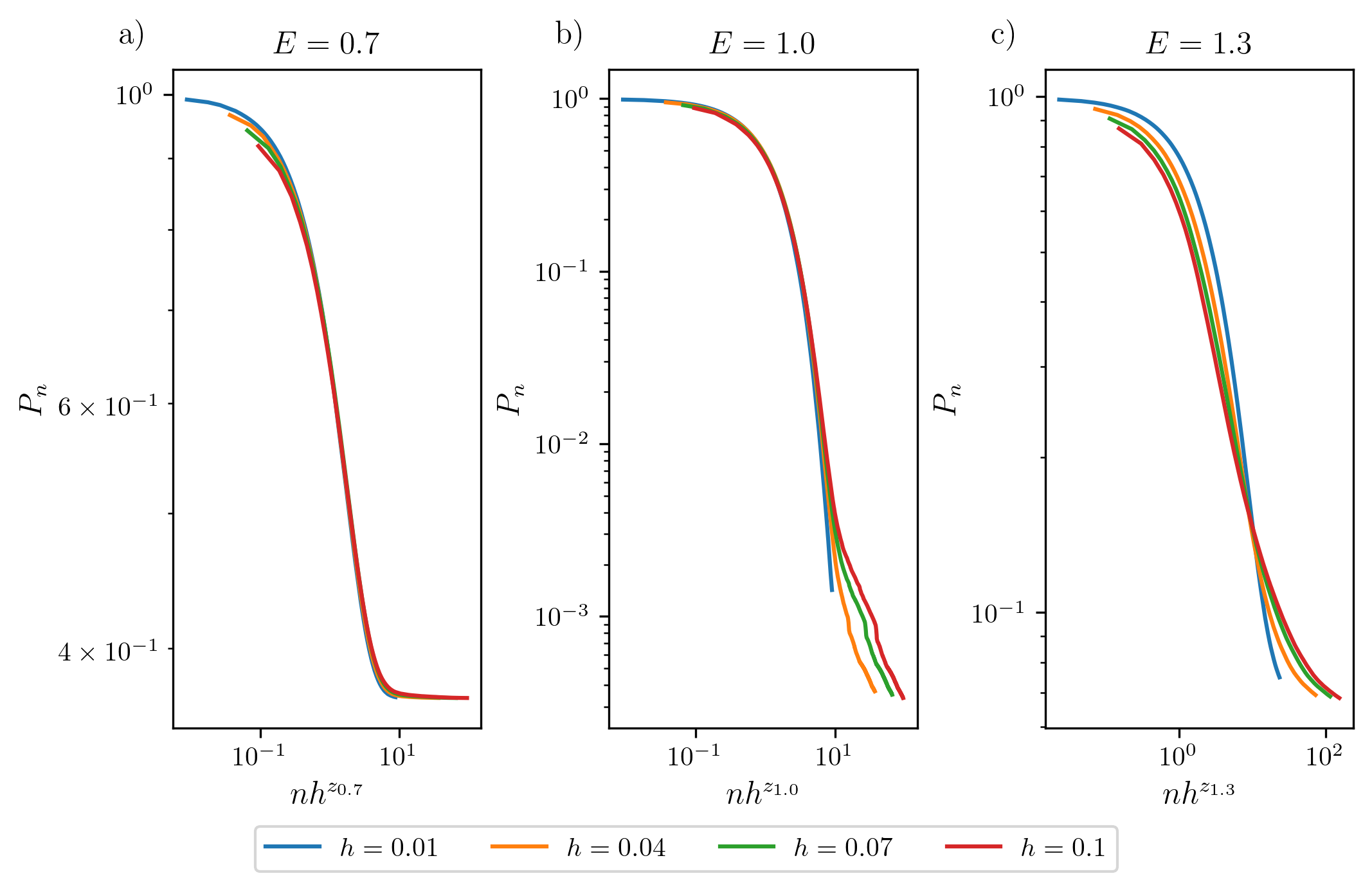}
    \caption{Universal decay curves obtained by rescaling the collision number axis according to the transformation $n \rightarrow n' = n h^z$. The panels correspond to energies (a) $E=0.7$, (b) $E=1.0$, and (c) $E=1.3$, using the respective scaling exponents $z$ found in Fig.~\ref{fig:zfit}. The excellent collapse of the data for different hole sizes onto a single curve demonstrates the scaling invariance of the escape dynamics.}
    
    \label{fig:universal}
\end{figure}

The discovery of these scaling exponents allows for the collapse of the individual decay curves onto a universal plot by rescaling the number of collisions, $n$, according to the transformation $n' = n \cdot h^z$. When $P(n)$ is plotted against this new variable $n'$, the data for all hole sizes align onto a single master curve, confirming the universality of the escape dynamics Fig~\ref{fig:universal}.

\section{Conclusions}
\label{sec:conclusions}

In this work, we have conducted a detailed numerical investigation of the escape dynamics of a particle in a circular billiard under the influence of a uniform gravitational field. By introducing two symmetric holes on the boundary, we created an leaky Hamiltonian system and systematically explored how its phase space structure and escape characteristics depend on two key parameters: the particle total energy, $E$, and the size of the escape holes, $h$.

Our analysis revealed a complex dynamical landscape. The most significant finding is the clear evidence of a transition from a strongly chaotic regime at low energies to a non-hyperbolic regime with a mixed phase space for energies $E \gtrsim 1.0$. This transition is physically marked by the emergence of large Kolmogorov-Arnold-Moser (KAM) islands, which introduce regions of stability and trapping into the otherwise chaotic dynamics.

We have demonstrated that quantitative measures from nonlinear dynamics serve as excellent indicators of this transition. The basin entropy, $S_b$, effectively captures the change in the system predictability, showing a general increase toward a peak near the transition energy, followed by fluctuations in the mixed-phase-space regime. Similarly, the mean escape time, $\bar{\tau}$, provides a clear temporal signature of the transition, with a sharp increase caused by the "stickiness" of the newly formed KAM islands. Our analysis of the Survival Probability, $P(n)$, reinforces these conclusions, showing a clear shift from pure exponential decay in the fully chaotic regime to a slower, power-law-like decay that saturates at a constant value in the mixed regime, directly quantifying the fraction of permanently trapped orbits.

\section*{Acknowledgments}

This work has been supported by grants from the Brazilian Government Agencies CNPq and CAPES. P. Haerter received partial financial support from the following Brazilian government agencies: CNPq (140920/2022-6), CAPES (88887.898818/2023-00). A. F. Bosio thanks the Coordenação de Aperfeiçoamento de Pessoal de Nível Superior – Brasil (CAPES) – Finance Code 001 and the Brazilian Federal Agency (CNPq) under Grant Nos. 403120/2021-7, 301019/2019-3. R. L. Viana received partial financial support from the following Brazilian government agencies: CNPq (403120/2021-7, 301019/2019-3), CAPES (88881.143103/2017-01). E.D.L. acknowledges support from Brazilian agencies CNPq (No. 301318/2019-0, 304398/2023-3) and FAPESP (No. 2019/14038-6 and No. 2021/09519-5). M.A.F. Sanjuán  acknowledges financial support from the Spanish State Research Agency (AEI) and the European Regional Development Fund (ERDF, EU) under Project No. PID2023-148160NB-I00 (MCIN/AEI/10.13039/ 501100011033).

\bibliographystyle{elsarticle-num}
\bibliography{Portal}
\end{document}